\documentstyle[11pt,IAUS215,twoside,epsf]{article}

\def\edcomment#1{\iffalse\marginpar{\raggedright\sl#1\/}\else\relax\fi}
\reversemarginpar

\begin{document}
\vspace*{1cm}
\title{Dynamical shear instability}
 \author{Raphael Hirschi, Andr\'e Maeder \& Georges Meynet}
\affil{Geneva Observatory, CH--1290 Sauverny, Switzerland}
%\author{Andr\'e Maeder \& Georges Meynet}
%\affil{Geneva Observatory, 
%CH-1290 Sauverny, Switzerland}

\begin{abstract}
The dynamical shear instability is an important mixing process in the advanced stages of the evolution of
massive stars. We calculated different 
models of 15 $M_{\odot}$ with an initial rotational velocity, $\upsilon_{\rm{ini}}= 300$ km/s 
to investigate its efficiency. 
We found that the dynamical shear instability has a timescale shorter than Oxygen burning
timescale and that it slightly enlarges the 
convective zones and smoothens the omega gradients throughout the evolution.
 However, its effect is too localized to slow down the core of the star. \\
\end{abstract}

\section{Theory \& computer model}

The stability criterion for dynamical shear instability is the 
Richardson criterion:
$Ri=\frac{N^2}{(\partial U/\partial z)^2} > \frac{1}{4}=Ri_{\rm{c}}$,
where 
$U$ is the horizontal velocity, $z$ the vertical coordinate and $N^2$
the Brunt-V\"ais\"al\"a frequency. Using this criterion, the timescale of the instability is 
the dynamical timescale and the instability is called dynamical shear instability (Endal \&
Sofia 1978). If heat
losses are accounted for in $Ri$ (Maeder 1997), the timescale is the thermal timescale and in that case the 
instability is called secular shear instability.
The critical value, $Ri_{\rm{c}}=1/4$,  is
used by most authors as the limit for the occurrence of the dynamical shear.
However, recent studies (Canuto 2002; Br\"uggen \& Hillebrandt 2001) show that turbulence may occur as
long as $Ri<\ \sim 1$. \\

Different formulae for the corresponding diffusion coefficient, $D$,
are used at the present time (Zahn 1992; Maeder 1997; Heger et al 2000;
 Br\"uggen \& Hillebrandt 2001).
The following dynamical shear diffusion coefficient suggested
 by J.-P. Zahn is used in this study:

\begin{equation}
D=\frac{1}{3}vl
=\frac{1}{3}\ \frac{v}{l}\ l^2
=\frac{1}{3}\ r\frac{\mathrm{d}\,\Omega}{\mathrm{d}\,r} \ \Delta r^2
=\frac{1}{3}\ r\Delta\Omega\ \Delta r
\end{equation}where $r$ is the mean radius of the zone where the instability occurs, 
$\Delta\Omega$ is the variation of $\Omega$ over the zone and 
$\Delta r$ is the extent of the zone. The zone is the reunion of consecutive shells
where $Ri< Ri_{\rm{c}}$. This is valid if the Peclet number, $P_e>1$. 
\\

The computer model used is the Geneva evolutionary code. Modifications (to be
described in a future paper) have been made to 
study the advanced stages of the evolution of massive stars. 
The models are calculated for 3 stars of
15 $M_{\odot}$ at $Z_{\odot}$ with $\upsilon_{\rm{ini}}=\ 300$ km/s. 
Convective stability is determined by the Schwarzschild criterion. 
The overshooting parameter, $d_{\rm{over}}/H_{\rm{P}}=0.1$ for H-- and He--burning and 0 afterwards. 
In the first model, the dynamical shear was not included. Then one model
was calculated with $Ri_c=1/4$ and one with $Ri_c=1$. The calculations reached shell
O--burning.   
Note that the computer model includes secular shear and meridional circulation for
any calculation.
\section{Results \& discussion}

The characteristic timescale of the dynamical shear is very short (a fraction of a
year) throughout 
evolution when using
 equation (1). We obtain diffusion coefficients between $10^{12}$
and $10^{14}$ cm$^2$/s. This is usually 1 or 2 orders of magnitude larger than 
the dynamical shear diffusion coefficients of Br\"uggen \& Hillebrandt (2001) or Heger et al (2000). 

However, the extent of the unstable zone is very small, a few $10^{-3}\ M_{\odot}$ 
so the shear mainly smoothens
the sharp $\Omega$--gradients but does not transport 
angular momentum over long distances. The total angular momentum of our models, at the
beginning and at the end of the calculations are of the same order of magnitude as
the values obtained by Heger et al (2000) in their models E15 and E15B.

Although dynamical shear strengthens convection, especially the
He--burning convective shell, 
the structure and the convective
zones are similar between the model without dynamical shear and the one 
with dynamical shear and
$Ri_{\rm{c}}=1/4$. Concerning the Richardson criterion, we see that there is a difference during 
Ne--burning between the model using $Ri_{\rm{c}}=1/4$ and $Ri_{\rm{c}}=1$. Indeed, using the latter, 
there is another large C--burning convective shell while the central convective core is
smaller. One could argue that our
diffusion coefficient is too high but we obtained a similar trend using the diffusion 
coefficient of Br\"uggen \& Hillebrandt (2001) with $Ri_{\rm{c}}=1$. 

The well known discrepancy between the observed and predicted pulsar rotation periods might be 
reduced with the interaction between rotation and magnetic
field (see contributions by Heger and Spruit in this volume) or 
with a formula linking dynamical and secular shear in the advanced stages of
the evolution of massive stars.
\\

\noindent Poster copy @ http://obswww.unige.ch/$\sim$hirschi/work/cancun02.ps

\end{document}